# Measurement of the cross section for the reaction $^{127}$I$(\nu_e, e^-)^{127}$Xe$_{\text{bound states}}$ with neutrinos from the decay of stopped muons

J. R. Distel,[*] B. T. Cleveland,[†] K. Lande, C. K. Lee, and P. S. Wildenhain

*Department of Physics and Astronomy, University of Pennsylvania, Philadelphia, Pennsylvania 19104, USA*

G. E. Allen[‡] and R. L. Burman

*Los Alamos National Laboratory, Los Alamos, New Mexico 87545, USA*

The cross section for the reaction $^{127}$I$(\nu_e, e^-)^{127}$Xe$_{\text{bound states}}$ has been measured for electron neutrinos from the decay of stopped muons to be $[2.84 \pm 0.91 \text{ (stat)} \pm 0.25 \text{ (syst)}] \times 10^{-40}$ cm$^2$. A tank containing 1540 kg of $^{127}$I in the form of NaI solution was placed 8.53 m from the Los Alamos Meson Physics Facility beamstop where it received a typical flux of $5 \times 10^7 \nu_e/$(cm$^2$ s). The $^{127}$Xe atoms produced by neutrino capture were extracted from the target solution, placed in miniature proportional counters, and their number was determined by counting. This is the first measurement of a neutrino capture cross section for an I nucleus and is in good agreement with a recent calculation.

## I. INTRODUCTION

For nearly 40 years physicists have been going to great lengths (and depths) to study the neutrinos emitted by nuclear fusion reactions in the sun. This work began with the pioneering effort of Raymond Davis, Jr., the Homestake chlorine experiment [1], whose major goal was to verify that nuclear fusion was taking place in the sun by observation of neutrinos, primarily those from the decay of $^7$Be and $^8$B. After taking data for 25 years, this experiment found only 2.56 ± 0.23 SNU [2], about 1/3 of the value predicted by the present standard solar model of $7.6^{+1.3}_{-1.1}$ SNU [3], where 1 SNU is defined as 1 interaction/s in a target that contains $10^{36}$ atoms of the neutrino absorbing isotope. Subsequent measurements of the solar neutrino capture rate with a gallium target, which is sensitive to the lower-energy $pp$ neutrinos [4], and of the $^8$B flux with a water target [5], also observed a solar neutrino flux that was less than predicted. For more than 30 years the cause of this difference between measured and expected neutrino signal was not understood and it became known as the "solar neutrino problem".

Recent measurements at the Sudbury Neutrino Observatory [6], which uses a $^2$H target, strongly support the interpretation of neutrino oscillations as the cause of the reduced solar neutrino flux. The agreement between the total $^8$B neutrino flux measured by the neutral-current reaction (which has equal sensitivity to all active neutrino flavors) with the predictions of the standard solar model, imply that a major fraction of the solar $\nu_e$ neutrinos oscillate into $\nu_\mu$ and/or $\nu_\tau$ neutrinos. Measurements by KamLAND [7] of the $\bar{\nu}_e$ flux from distant nuclear reactors further strengthen the oscillation interpretation.

Haxton pointed out 15 years ago that $^{127}$I would make an attractive solar neutrino experiment [8]. Neutrinos would be detected by the reaction $^{127}$I$(\nu_e, e^-)^{127}$Xe, which has an effective threshold of 789 keV, thus giving sensitivity to both intermediate-energy ($^7$Be, $pep$, carbon-nitrogen-oxygen cycle) and high-energy ($^8$B) solar neutrinos. Since the target would be an I-containing liquid in a tank and Xe would be extracted by a circulating gas flow, then purified and counted in a small proportional counter, an I experiment would in many ways be similar to the Homestake chlorine experiment.

Although an I experiment has many advantages, such as 100% isotopic abundance, a very favorable counting scheme, and a high Coulomb barrier (which gives low sensitivity to background from local protons and $\alpha$ particles), it suffers from the disadvantage that neutrino capture can only proceed to excited states. As a consequence, although guidance can be obtained from theoretical calculations [8–10] and from measurements of the $(p, n)$ reaction in the forward direction at high proton energies, the neutrino capture cross section of I for the various solar neutrino components must ultimately be determined by direct measurements with neutrinos. This situation is in contrast to the other radiochemical solar neutrino experiments, $^{37}$Cl, for which the relevant cross sections can be inferred from measurements of the decay of the mirror nucleus $^{37}$K, and $^{71}$Ga, for which the capture rate is dominated by transitions to the ground state of $^{71}$Ge.

This paper describes a first step in such a calibration of an I solar neutrino detector. It is a measurement at the Los Alamos Meson Physics Facility (LAMPF) of the capture cross section of $^{127}$I for $\nu_e$ from the decay of stopped muons. The purpose was to check on the calculations of the high-energy response of I. Further, this observation of the $^{127}$I$(\nu_e, e^-)^{127}$Xe$_{\text{bound states}}$ reaction is the first reported cross section measurement of a neutrino reaction on a nucleus heavier than $^{56}$Fe [11].

We describe the experimental technique in Sec. II. Data analysis and the experimental results are given in Sec. III, with a summary and conclusions in Sec. IV.

---

[*]Present address: Los Alamos National Laboratory, Los Alamos, NM 87545.
[†]Present address: Sudbury Neutrino Observatory, Lively, Ontario P3Y 1M3, Canada.
[‡]Present address: Center for Space Research, MIT, Cambridge, MA 02139.

## II. EXPERIMENTAL DESIGN

### A. Overview

This experiment measured the rate of the reaction $^{127}$I$(\nu_e, e^-)^{127}$Xe$_{\text{bound states}}$ using the $\nu_e$ flux from the decay of stopped muons at the LAMPF beamstop. By definition of the cross section, the production rate $p_{\text{beam}}(t)$ of Xe by beam-associated neutrinos is given by

$$p_{\text{beam}}(t) = N_I \Phi_\nu(t) \overline{\sigma}_\nu, \quad (1)$$

where $N_I$ is the number of $^{127}$I target atoms, $\Phi_\nu(t)$ is the time-dependent $\nu_e$ flux whose normalized spectral shape is $S_\nu(E)$, and $\overline{\sigma}_\nu = \int S_\nu(E) \sigma(E) dE$ is the flux-shape weighted cross section, whose measurement we report here.

Techniques similar to those used in the Homestake chlorine solar neutrino experiment were employed to extract and then to detect the $^{127}$Xe atoms. The I target was a large volume of NaI solution contained in a tank that was instrumented with a pump and plumbing that enabled the Xe atoms to be swept from the liquid and collected. Extractions were performed at several week intervals. The sample from each extraction was purified and the collected Xe atoms were placed in a small proportional counter. All events from this counter were recorded, typically for a period of a year, at the end of which the counting data was searched for the characteristic decay signature of $^{127}$Xe back to $^{127}$I which occurs with a half-life of 36.4 d. By combining the number of Xe events seen in the counter with the measured values for counting and extraction efficiency, the $^{127}$Xe production rate in the tank could be calculated. In addition to $^{127}$Xe production from the desired neutrino capture process, competing background reactions also contributed.

Discussion of the beamstop and neutrino source is in Sec. II B. A description of both primary shielding elements and secondary shielding components can be found in Sec. II C. Details of the tank assembly, target material, and extraction apparatus are given in Sec. II D. Extraction parameters and efficiencies are discussed in Sec. II E, counting procedures in Sec. II F, and counting efficiency in Sec. II G.

### B. Neutrino source

The LAMPF beamstop facility provided a calibrated, high-intensity source of neutrinos, well suited to a total cross section measurement. Protons of 800-MeV kinetic energy produce pions in the beamstop. The majority of the $\pi^+$ come to rest in the beamstop where they decay into $\mu^+$ and $\nu_\mu$. The subsequent decay of the stopped muon gives $\overline{\nu}_\mu$ and $\nu_e$. The energy spectra of these three neutrinos are shown in Fig. 1. The $\nu_e$ spectrum is the Michel spectrum of muon decay at rest. Only the $\nu_e$ neutrinos are of interest to us because the energy of the $\nu_\mu$ neutrinos is below threshold for charged-current reactions and because neutral-current reactions cannot produce the I to Xe transition.

To compute neutrino production at the beamstop facility the neutrino fluxes from $\pi^+$ and $\mu^+$ decay at rest were calculated with a Monte Carlo computer program designed for spallation targets and beamstop facilities at medium-energy proton accelerators. A detailed description of the code is available in Ref. [12] and so only a brief outline is given here. The program uses proton reaction cross sections, pion production and absorption cross sections, and particle transport to calculate the neutrino fluxes from the decays of positive pions and of positive and negative muons. The proton beam is transported, with energy loss, through the beamstop facility geometry. At a Monte Carlo chosen proton interaction point, positive and negative pions, weighted by the production cross sections, are selected with initial energy and angle according to measured cross sections. As the pions are tracked through the geometry they are allowed to inelastically scatter, to multiple-Coulomb scatter, to be absorbed, or to decay. Absolute normalization was provided by measurements [13] made on an instrumented mockup of a simplified beamstop; the event-by-event production of pions, followed by signals from pion and muon decay, was used to infer the rate of stopped $\pi^+$ production per incident proton. As input to the code, the LAMPF beamstop facility was modeled in sufficient detail to reproduce an initial target of water in an aluminum container, a number of isotope production targets primarily consisting of aluminum boxes, and finally the proton beamstop composed of water-cooled copper disks.

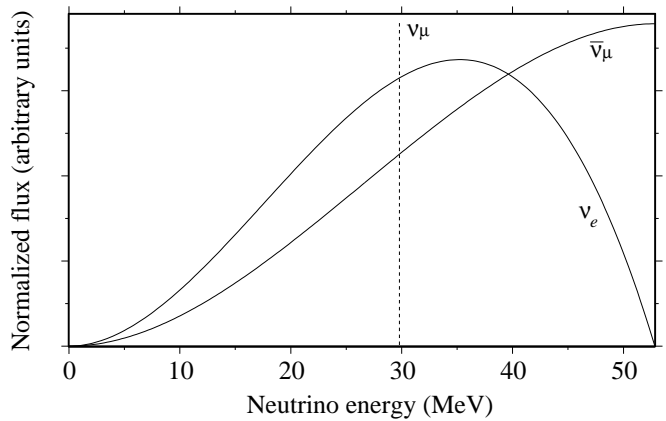

FIG. 1: Energy spectra of the $\nu_\mu$ neutrino from $\pi^+$ decay at rest, and of the $\overline{\nu}_\mu$ and $\nu_e$ neutrinos from $\mu^+$ decay at rest. Only the $\nu_e$ neutrinos can produce transitions from I to Xe.

From the Monte Carlo simulation of $\pi^+$ and $\mu^+$ decays at rest in the beamstop facility, the source region was approximately localized within a cylinder of less than 100 cm length by 25 cm radius. Because these decays occur at rest, an isotropic distribution of neutrinos results. The $\nu_e$ flux could be inferred from the number of $\mu^+$ decays at rest, and expressed as the mean number of $\nu_e$ per incident proton. This number, multiplied by the beam current, gives the $\nu_e$ intensity. During the two running seasons of this experiment $\nu_e$/proton varied, due to changes in the isotope production targets, from 0.082 to 0.092. As described in detail in Ref. [12], the absolute error on the $\nu_e$ flux at the beamstop target is 7.3%.

The typical $\nu_e$ flux at the target tank can be calculated from the proton current, the $\nu_e$/proton, and the source-to-tank distance. For a distance from the tank center to the beamstop of 8.53 m, and for the nominal proton current of 0.8 mA, the typical running period had a neutrino flux at the middle of the tank of $5 \times 10^7 \nu_e/(\text{cm}^2 \text{ s})$.

### C. Shielding

To reduce the flux of cosmic-ray particles and beam-associated neutrons, the iodine tank assembly was placed in a well-shielded room at the neutrino beamstop area. This room was constructed with thick concrete and steel ceiling and walls; in addition, secondary water shielding was erected above and around the tank to attenuate the flux of low-energy neutrons reaching the detector. To maximize neutrino flux, the tank was placed as near to the neutrino source as allowed by shielding constraints.

The overhead shielding in the roof consisted of at least 1.52 m of steel plus 1.83 m of concrete for a thickness of 1750 g/cm$^2$. The north wall, between the beamstop and the detector, contained 6.7 m of steel, cast iron, and lead. The side walls contained at least 1 m of steel and the floor at least 1 m of concrete. The additional shielding for the attenuation of low-energy, beam-associated neutrons consisted of water-filled containers that were placed around all sides, and above and below the tank. A water blanket at least 60-cm thick was thus provided about all six tank faces.

### D. Tank

The iodine tank assembly, diagrammed in Fig. 2, consisted of a rectangular steel tank, a magnetic-drive circulation pump, a liquid circulation loop, a gas circulation loop, and a gas-liquid mixing device or eductor. Circulation was established by the pump which drove liquid around the liquid loop through the eductor. The eductor was a conical nozzle in which the increased liquid velocity resulted in a decreased gas pressure, thereby pulling gas from the gas line into the nozzle region where it mixed with the liquid. The liquid flow thus established the gas flow and no gas pump was needed. Extraction traps were placed in the gas loop to separate the Xe gas from the circulating He gas. The gas line carried gas from the tank to this extraction system, and then returned the gas through the eductor and liquid line. To prevent the inleakage of atmospheric air the entire gas loop, including the tank, was a sealed system.

The detector tank was a 1.8 m × 1.8 m × 0.8 m rectangular vessel made from 6.4-mm-thick steel sheets. To increase structural rigidity 2.5-cm-diameter tubes were used as tie rods, which connected the two large faces of the tank together on a 30 cm by 30 cm checkerboard grid. An additional two tubes had 5-cm-diameter hollow centers, sufficient to accommodate a neutron source.

The detector tank contained 2220 ± 60 liters of NaI solution which was 50.8% NaI by weight and had a density of 1.614 ± 0.007 g/cm$^3$. Since iodine is monoisotopic, the $^{127}$I content in the tank was 1540 ± 46 kg or $(7.31 \pm 0.22) \times 10^{27}$ atoms. The volume of the tank not occupied by NaI solution was ≈250 liters and contained He gas at a pressure of ≈5 psi.

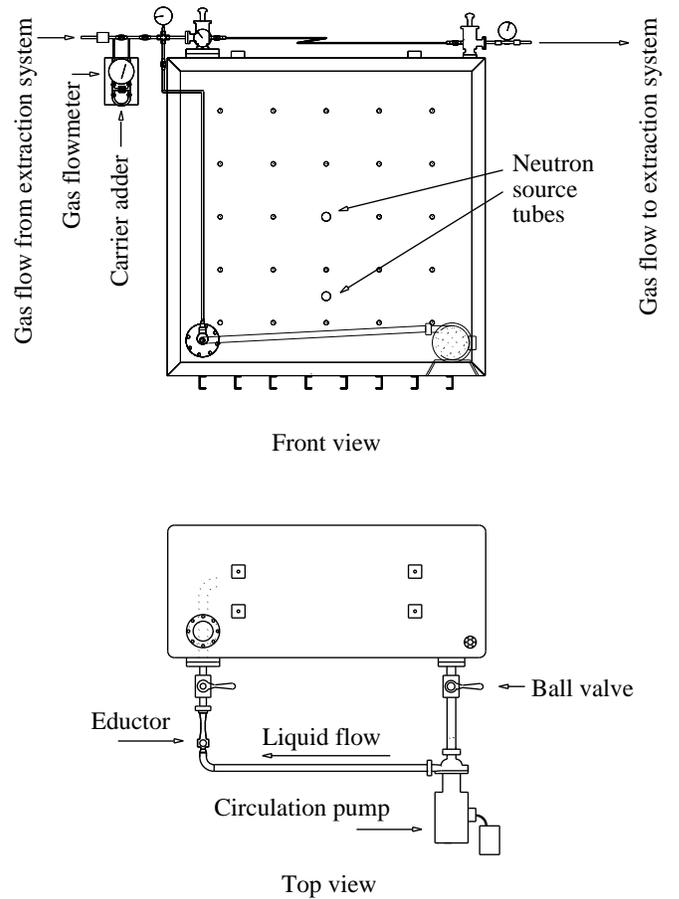

FIG. 2: Schematic diagram of tank, liquid line, and gas line.

The carrier adder assembly shown in Fig. 2 was a system of valves that allowed the return flow from the extraction system to be diverted into a vessel containing a small well-measured volume (≈0.1 standard cm$^3$) of normal Xe gas. At the beginning of each run this Xe gas was swept from the carrier adder into the main return line and ultimately into the tank, where it remained during the exposure to neutrinos. This added Xe was used to measure the extraction efficiency, as will now be described.

### E. Extraction and gas purification

Xe atoms were removed from the gas flow by an extraction system containing a molecular sieve column to remove water vapor and a main extraction trap filled with ≈3 kg of low-background charcoal. The main trap was immersed in liquid nitrogen during extractions; it stopped the Xe atoms but allowed the He carrier gas to pass through. For each extraction a circulatory gas flow from the tank through the trap and back to the tank was maintained at a fixed flow rate of ≈70 l/m for

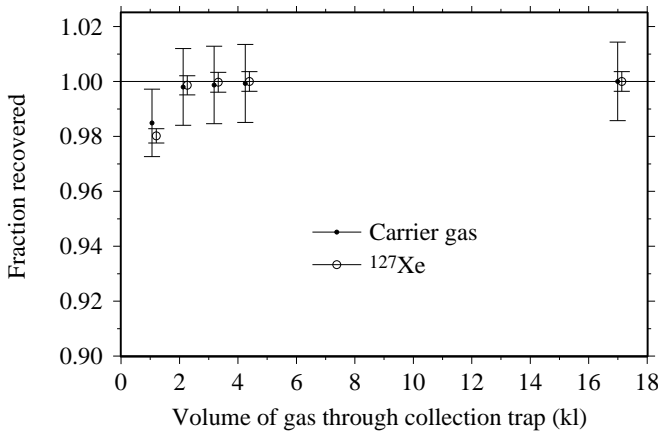

FIG. 3: Fraction of carrier and $^{127}$Xe recovered as a function of volume of He gas that flowed in differential neutron source runs.

4 h. At the end of this time, the collected sample was transferred from the main charcoal trap to a small secondary trap for transport from Los Alamos to Philadelphia, where further processing and counting were carried out, as described below.

One expects the volume of Xe gas extracted $V(t)$ to be an exponential function of the volume of sweep gas $G(t)$ sent through the extraction system

$$V(t) = V_0(1 - e^{-CG(t)}),  \quad (2)$$

where $V_0$ is the initial volume of Xe and $C$ is an extraction coefficient. The extraction efficiency $\varepsilon_E$ is defined as $V(t)/V_0$. It is determined in practice as the ratio of the measured volume of extracted Xe gas to the volume of added carrier.

There are several subtle points in the assumption that a 0.1 cm$^3$ aliquot of gas ($10^{17}$ atoms) introduced into the gas phase correctly mimics the extraction properties of several hundred atoms produced locally, inside a liquid via a neutrino capture process. To check this assumption and prove that normal Xe carrier gas was extracted in the same way as $^{127}$Xe produced *in situ*, an experiment was conducted with a PuBe neutron source inserted into one of the central 5-cm tank tubes. Neutrons from the source produce $^{127}$Xe through the two-step process of $(n, p)$ scattering, followed by $^{127}$I$(p,n)^{127}$Xe. A source of intensity $2.2 \times 10^6$ n/s, inserted for a 4.8-d exposure, produced $\approx 3 \times 10^5$ $^{127}$Xe atoms in the target. At the end of exposure, a series of short duration, differential extractions were performed. By comparing the recovered volume fraction of carrier to the recovered number fraction of $^{127}$Xe in each of these short extractions, a quantitative statement about the extraction behavior of carrier versus locally produced $^{127}$Xe can be made.

Figure 3 shows the results of this comparison for the five differential extraction runs. Each of the first four extractions was 15 min in length. The final extraction was for 3 h so the total extraction time of 4 h equaled the duration of a normal extraction run. Two things can be noted from Fig. 3: First, with 98.5% of the inserted carrier being recovered in the first 15 min sample, the data imply a very rapid extraction rate, a "1/e" time of about 2 min. This suggests that the normal extraction time of 4 h has an extremely large safety factor.

The second feature of Fig. 3 is the good agreement between the extraction efficiency as measured with carrier gas and with $^{127}$Xe. This provides reasonable assurance that the fractional volume of extracted carrier correctly estimates the extraction efficiency for locally-produced $^{127}$Xe. We performed additional sweeps several weeks after the high-activity, differential runs. No additional $^{127}$Xe was detected, thus ruling out any long term trapping processes for $^{127}$Xe in the NaI solution.

Processing of the small secondary trap which contained the gas collected from the detector tank required a two-step gas purification process. Gettering with hot Ti dissociated and chemically bound air gases ($N_2$, $O_2$, $CO_2$, $H_2O$) while transmitting the noble gases (He, Ne, Ar, Kr, Xe, Rn). Next, a chromatography column was used to separate Xe from Ar, Kr, and Rn. After chromatography, the Xe sample was placed in a proportional counter, and a small amount of P-10 gas (90% Ar, 10% CH$_4$) added to bring the internal pressure of the counter to $\approx$1.0 atm. The Xe efficiency of these purification steps was approximately 100%.

### F. Counting

After extraction and processing, the proportional counters containing the Xe samples were installed in a 20–30 cm thick Pb and Cu shield and were counted for 8–10 half-lives ($t_{\text{half}} = 36.4$ d) to determine the number of $^{127}$Xe atoms present. $^{127}$Xe decays by orbital electron capture back to $^{127}$I with the characteristic signature of at least one Auger electron in coincidence with at least one nuclear $\gamma$ ray. This coincidence was exploited by placing the miniature proportional counter, capable of detecting the Auger emission, inside a NaI crystal, capable of detecting the $\gamma$ radiation.

Of the $^{127}$Xe decays by orbital electron capture, 83.5% occur from the $K$ shell, 13.0% from the $L$ shell, and 2.9% from the $M$ shell. The vacancy created by capture is refilled by a higher order orbital electron resulting in the emission of at least one Auger electron; 67.4% of all decays produce Auger electrons of energy near 4.7 keV. In addition to the radiation from atomic shell rearrangements, the decay of $^{127}$Xe also produces $\gamma$ rays from nuclear deexcitations. Because the g.s. to g.s. transition between $^{127}$Xe (1/2$^+$) and $^{127}$I (5/2$^+$) is not allowed, the decay proceeds to an excited state in the $^{127}$I nucleus, roughly 50% to the 203-keV level and 50% to the 375-keV level. Essentially all the 203-keV level decays go directly to the $^{127}$I ground state, with the emission of a 203-keV $\gamma$ ray. From the 375-keV level, roughly half the decays go directly to the ground state with the emission of a 375-keV $\gamma$ ray; the other half emit a 172-keV $\gamma$ ray and jump to the 203-keV level.

The performance of the counting system is best demonstrated with data from a strong $^{127}$Xe source. The coincident counts from such a source, for both Auger electrons and $\gamma$-ray emissions, are shown as a two-dimensional plot in Fig. 4.

The miniature proportional counters used in this experiment were cylindrical in design with a single anode wire running along the axis of the iron cathode, and were constructed of ultrapure materials. Cathodes had a length of 25–30 mm, an outside diameter of 6 mm, and an inside diameter of 5 mm. The counter capacitance was $\approx 0.3$ pF. When filled with P-10 counting gas, the critical field was about $3 \times 10^6$ V/m.

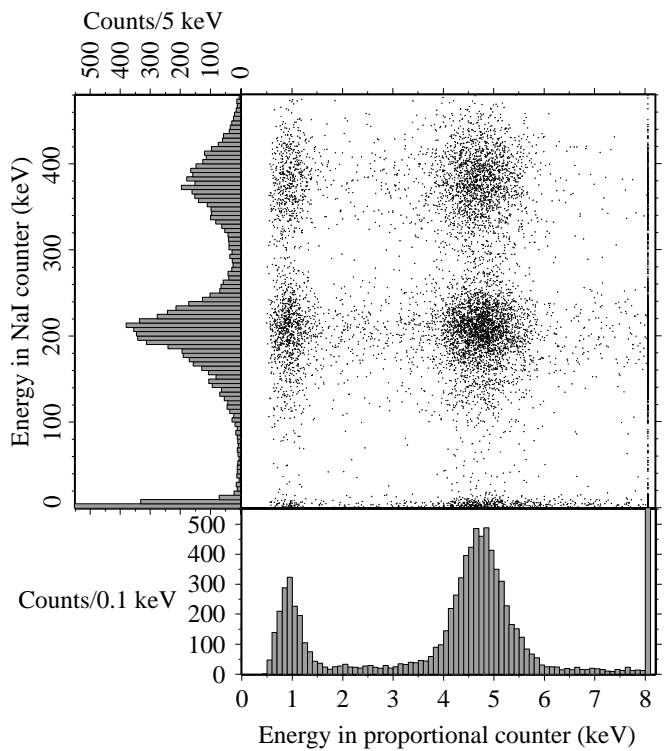

FIG. 4: Measured Auger electron and nuclear $\gamma$-ray coincidence spectra from the decay of $^{127}$Xe in a proportional counter inside the well of a NaI detector. The spectrum of Auger electrons from the $^{127}$Xe $\beta$ decay is at the bottom and the spectrum of nuclear $\gamma$ rays from the subsequent $^{127}$I deexcitation is at the left. The electronics was triggered by the proportional counter with a threshold at 0.53 keV and 10 000 events are shown. The peaks in the Auger spectrum are the $M$ peak at 0.9 keV, the $L$ peak at 4.7 keV, and electronics saturation at 8.1 keV. The prominent peaks in the NaI spectrum are from 203-keV and 375-keV $\gamma$ rays. The events at low energy in the NaI spectrum occurred when no $\gamma$ rays interacted in the NaI detector, mainly because they escaped through its well.

A voltage of 1000–1200 V was applied to the counter cathode. Because multiplication occurs only within a region of radius of $\approx 0.1$ mm from the center of the anode, i.e., about 1/500 of the active volume, essentially all decays occurred outside the gas multiplication region. As the mean energy to produce an electron-ion pair in Ar is about 26 eV, the Auger electrons of energy $\approx 5$ keV produced a few hundred electron-ion pairs. For a counter of the geometry outlined above typical gas multiplication factors were about 3000 with total charge of order $10^{-13}$ C.

The signal from the anode was direct coupled to a charge-sensitive preamplifier with sensitivity of 1 V/pC and rise time of 50 ns. The preamplifier output signal was split and sent to two amplifiers. The first, a standard shaping amplifier, measured the integrated charge or energy of the pulse. The second, a timing filter amplifier, provided a measure of the pulse rise time, called the ADP (amplitude of differentiated pulse). This ADP signal was proportional to the energy of the pulse and inversely proportional to its rise time. The ADP was used to distinguish localized, point-ionization events (such as those from an Auger electron), whose ADP was large, from extended-ionization tracks (such as those from a through-going electron arising from Compton scattering), whose ADP was small.

During the counting period of 10–12 months, each counter was placed inside the well of a NaI(Tl) crystal, with four counters sharing one crystal. A copper electrostatic shield surrounded the four counters and their associated preamplifiers. To attenuate local $\gamma$ rays the entire assembly of NaI crystal and electrostatic shield was enclosed within 25–35 cm of Pb and steel.

For every event from the proportional counter the energy, ADP, any coincident NaI signal, and the date and time were recorded. Calibrations were made with the 5.9-keV Mn x ray from a small encapsulated $^{55}$Fe source positioned close to the thin window of the proportional counter. They were performed at the start of counting, at the end of counting, and periodically every 60 days in between. During the initial calibration the counter high voltage and the amplifier gains were adjusted to provide maximum sensitivity to the $L$-Auger electrons (4.7 keV) that constituted the dominant signal. The $^{55}$Fe calibration was quite suitable in locating the $L$-peak energy region as it gave two peaks (at 1.6 keV and 5.9 keV) which straddled the 4.7-keV region. Calibrations were also performed on the NaI crystal with a $^{137}$Cs source.

### G. Measurement of counting efficiency

Accurate determination of the proportional counter Xe detection efficiency was required to convert from the number of observed $^{127}$Xe decays to the total number of $^{127}$Xe atoms initially present in the counter. The efficiencies of both the gas proportional counters and the NaI crystals were measured by utilizing the coincidence between the nuclear $\gamma$ rays and the Auger electrons.

The counter to be calibrated was filled with a hot sample of Xe and placed inside a NaI crystal. The data-taking electronics was triggered either by a signal from the counter or from the crystal. For either trigger choice, measurements were made of the singles rate in the counter, the singles rate in the crystal, the background rates in the NaI crystals, and the coincident rate of counter and crystal. In some cases segmented NaI $\gamma$-ray detectors were used and it was also possible to use a stronger coincidence requirement of two $\gamma$ rays appearing in different segments of the detector. In both cases, the desired efficiencies could be directly computed from the measured background rates, singles rates, and coincidence rates. The volume efficiency of the proportional counters was 70–90%, and the efficiency of the NaI crystals was 80–90%.

## III. EVENT SELECTION, DATA ANALYSIS, AND RESULTS

### A. Overview

The Xe atoms produced during the exposure to the neutrino beam are unstable and decay at a rate governed by the decay constant $\lambda = \ln 2/t_{\text{half}}$. Defining the total Xe production rate as $p(t)$, the mean number of Xe atoms $N_{\text{tank}}$ present in the tank at the time of end of exposure $\theta$ is

$$N_{\text{tank}}(\theta) = \int_0^\theta p(t) e^{-\lambda(\theta-t)} dt, \quad (3)$$

where we have defined time zero at the beginning of exposure.

The production rate was the sum of two terms,

$$p(t) = p_{\text{bkgNB}} + p_{\text{beam}}(t), \quad (4)$$

where $p_{\text{bkgNB}}$ is the rate from time-independent background processes, such as cosmic rays, and $p_{\text{beam}}(t)$ is the rate from processes which had a dependence on the accelerator beam. As will be seen in Sec. III D, the only significant contributor to Xe production from the beam was neutrinos, given by Eq. (1).

In practice, the beam current and thus the neutrino flux $\Phi$ was not constant. To model these changes we set

$$\Phi(t) = \sum_{i=1}^l \phi_i U(t_b^i, t_e^i), \quad (5)$$

where $l$ is the number of flux intervals during the $i$th of which the flux was constant at the value $\phi_i$, $t_b^i$ and $t_e^i$ are the beginning and ending times of flux interval $i$, and $U$ is the Heaviside unit step function, 1 if $t_b^i < t < t_e^i$ and 0 otherwise.

Substituting Eqs. (4), (1), and (5) into Eq. (3), and carrying out the integration, we obtain

$$\lambda N_{\text{tank}}(\theta) = p_{\text{bkgNB}}(1 - e^{-\lambda\theta})$$
$$+ N_I \overline{\sigma}_\nu \sum_{i=1}^l \phi_i [e^{-\lambda(\theta - t_e^i)} - e^{-\lambda(\theta - t_b^i)}]. \quad (6)$$

This is the basic equation that will be used to determine the cross section.

The criteria used to select candidate Xe events is considered in Sec. III B. In Sec. III C the number of selected events is combined with information on the extraction efficiency, the counting efficiency, and exposure times to determine the total production rate for each run. Backgrounds are discussed in Sec. III D and the resulting cross section is given in Sec. III E.

### B. Event rates and event selection

For a typical counter there were about 2000 single counter triggers each day. In the 60-d interval between counter calibrations $\approx 10^5$ events were thus accumulated. The vast majority ($\approx 97\%$) of these events were due to cosmic rays which saturated the NaI crystal, depositing >800 keV, and were discarded. Of the remaining 3000 events, $\approx 85\%$ had low NaI energy and arose from local background processes. When these obvious background events were removed, some 400 events remained. A final preliminary cut removed events which saturated the proportional counter energy scale (>7 keV), produced either by $^{222}$Rn inside the counter or by the 33-keV $K$-Auger electron from $^{127}$Xe decay. This left $\approx 300$ events for a typical sample in the first interval of counting.

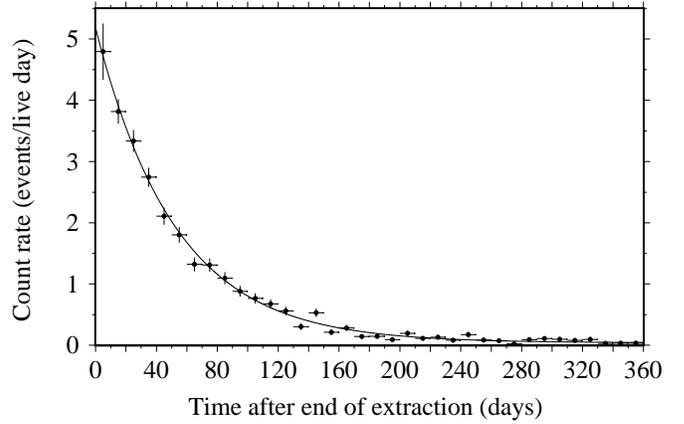

FIG. 5: Count rate for runs in Table I. The solid line is a fit to the data points with the 36.4-d half-life of $^{127}$Xe plus a constant background. The vertical error bar on each point is proportional to the square root of the number of counts and is shown only to give the scale of the error. The horizontal error bar is ±5 d, equal to the 10 d bin size.

Identification of coincident events in the Xe candidate population was achieved first, by requiring detection of one or more $\gamma$ rays of appropriate energy in the NaI crystal. A very broad window encompassing the expected 203-keV and 375-keV lines was used; this included virtually all (99%) of the $^{127}$Xe decays, but reduced the event sample by 20%. Guided by the $^{55}$Fe calibration performed at the beginning and end of the counting interval, events were then selected based on the energy deposited in the proportional counter. Since the energy of the $L$ peak in the $^{127}$Xe decay is 4.7 keV, a selection window of 3.7–5.7 keV was used, corresponding approximately to a 2 full width at half maximum interval. This reduced the number of events to about 140. A further cut, based on the ADP or rise time of the proportional counter pulse, as described in Sec. II F, typically removed another 20 events. The total number of events that remained is given in column 7 of Table I for 12 runs at LAMPF. The time distribution of these same events is shown in Fig. 5.

### C. Production rates

A maximum likelihood (ML) method [14] was used to determine the production rate and counter background rate for each run from the sequence of times of events that survived

TABLE I: Beam exposure data, counting data, and production rate for each run. The beam was off during runs 1 and 8 through 11 so these runs measured the background rate. Runs 3, 4, and 7 were counted under an overburden of ≈900 hg/cm² (1 hg = 10² g) in the Lehigh tunnel on the Northeast Extension of the Pennsylvania Turnpike; the other samples were measured in a basement location of the University of Pennsylvania physics building. The delay time is defined as the time from the end of extraction to the start of counting. The goodness of fit probability is calculated as described in Ref. [15] and has an accuracy of ±2% due to the finite number of simulations.

| Run ID | Exposure time (d) | Effective $\nu_e$ flux $F$ ($10^6$ $\nu$/(cm² s)) | Delay time (d) | Counting live time $T$ (d) | $\Delta$ | Number of counts Total | Number of counts $N_{Xe}$ | Goodness of fit probability (%) | Production rate ($^{127}$Xe atoms/d) |
|---|---|---|---|---|---|---|---|---|---|
| 1 | 24.80 | 0.0 | 14.6 | 205.0 | 0.635 | 167 | 167.0 | < 1 | 52.9 $^{+4.2}_{-4.0}$ |
| 2 | 10.78 | 33.56 | 26.5 | 328.6 | 0.576 | 96 | 75.2 | 93 | 42.0 $^{+5.9}_{-5.6}$ |
| 3 | 13.96 | 44.17 | 6.1 | 299.0 | 0.803 | 169 | 157.9 | 96 | 55.0 $^{+4.9}_{-4.7}$ |
| 4 | 14.99 | 40.94 | 9.3 | 265.4 | 0.804 | 138 | 134.4 | 50 | 56.6 $^{+5.4}_{-5.1}$ |
| 5 | 19.98 | 46.07 | 6.3 | 342.7 | 0.868 | 248 | 223.0 | 66 | 60.8 $^{+4.5}_{-4.3}$ |
| 6 | 20.98 | 39.71 | 6.2 | 280.7 | 0.862 | 226 | 206.5 | 84 | 50.4 $^{+4.1}_{-3.9}$ |
| 7 | 13.35 | 45.96 | 8.0 | 231.3 | 0.829 | 154 | 154.0 | 3 | 51.1 $^{+4.2}_{-4.2}$ |
| 8 | 36.54 | 0.0 | 7.5 | 254.4 | 0.717 | 257 | 249.2 | 35 | 48.5 $^{+3.4}_{-3.3}$ |
| 9 | 45.89 | 0.0 | 7.2 | 348.7 | 0.863 | 331 | 316.4 | 46 | 42.7 $^{+2.6}_{-2.5}$ |
| 10 | 22.73 | 0.0 | 7.3 | 337.8 | 0.815 | 240 | 208.7 | 12 | 43.9 $^{+3.4}_{-3.3}$ |
| 11 | 57.13 | 0.0 | 11.0 | 303.2 | 0.733 | 291 | 282.7 | 20 | 49.8 $^{+3.2}_{-3.2}$ |
| 12 | 28.02 | 40.97 | 9.0 | 330.5 | 0.831 | 334 | 318.6 | > 99 | 56.1 $^{+3.4}_{-3.3}$ |

all cuts. This was done by constructing a likelihood function

$$\mathcal{L}(a,b) = e^{-a\Delta/\lambda - bT} \prod_{i=1}^{n} (ae^{-\lambda(t_i - \theta)} + b), \quad (7)$$

and searching for the values of $a$ and $b$ that maximize $\mathcal{L}$. The variable $a$ is related to the total number of counts identified to be $^{127}$Xe decays $N_{Xe}$ by $a = \lambda N_{Xe}/\Delta$ and the variable $b$ is the background rate. In the likelihood function, $n$ is the total number of selected events, $t_i$ is the time of occurrence of Xe candidate event $i$, and $T$ is the live time of counting. The quantity $\Delta$ is the probability that the counter will be in operation at the time a Xe decay occurs. If the total number of counting intervals is called $k$ and the $j$th interval starts at time $t_{sc}^j$ and ends at time $t_{ec}^j$, then $\Delta = \sum_{j=1}^{k} \{\exp[-\lambda(t_{sc}^j - \theta)] - \exp[-\lambda(t_{ec}^j - \theta)]\}$. The one $\sigma$ error on production rate was obtained by finding the values of $a$ at which $\ln \mathcal{L}$ decreased from its value at the maximum by 0.5, where the background rate was chosen to maximize the likelihood function at these two points.

The values of $T$, $\Delta$, and $N_{Xe}$ for each run are given in Table I. A total of 2651 events were detected in the 12 runs and of these 2493.6 were ascribed to $^{127}$Xe. The number of background events was thus only 157.4, an average of one every 22.4 d. Proportional counter background was thus not a significant problem. The best fit half-life for all runs from the ML analysis was 35.7 ± 1.0 d, in agreement with the known value of 36.4 d. The probability that the observed sequence of events arose from the combination of $^{127}$Xe decay plus background events at a constant rate was calculated by the Monte Carlo method and is given in the next to last column of Table I. Some runs have rather low, and others rather high, probability of occurrence, but the distribution is entirely consistent with what is expected due to normal statistical variation.

The relationship between $N_{Xe}$, the number of counts identified to be $^{127}$Xe, and $N_{tank}$, the mean number of Xe atoms present in the tank at the end of exposure, is simply

$$N_{tank}(\theta) = \frac{N_{Xe}}{\varepsilon_E \varepsilon_C \Delta}. \quad (8)$$

For our measurements $\varepsilon_E$, the combined extraction and Xe processing efficiency discussed in Sec. II, had an average value of 0.907 ± 0.020. The counting efficiency $\varepsilon_C$ is the product of three efficiencies $\varepsilon_{Ctr}$, $\varepsilon_{NaI}$, and $\varepsilon_{ADP}$. $\varepsilon_{Ctr}$ refers to the proportional counter efficiency for counting Xe, and after all cuts was typically 0.43 ± 0.005. The two other factors $\varepsilon_{NaI}$ and $\varepsilon_{ADP}$ refer to the efficiency for detection of $\gamma$ rays by the NaI crystal and to the efficiency associated with the ADP cut, respectively. The product of the solid angle coverage (typically 0.950) and the $\gamma$ survival probability (0.901) yielded an average NaI efficiency of 0.850. A direct measurement of $\varepsilon_{ADP}$ was made by use of an intense Xe sample; the fraction of accepted events was 0.953.

If we combine Eq. (6) and (8) and rearrange terms we obtain

$$\frac{\lambda N_{Xe}}{\varepsilon_E \varepsilon_C (1 - e^{-\lambda \theta}) \Delta} = p_{bkg} N_B + N_I \overline{\sigma}_\nu F, \quad (9)$$

where we have defined $F$ to be the effective neutrino flux

$$F = \frac{1}{1 - e^{-\lambda \theta}} \sum_{i=1}^{l} \phi_i [e^{-\lambda(\theta - t_e^i)} - e^{-\lambda(\theta - t_b^i)}]. \quad (10)$$

The ML program directly calculated the left side of Eq. (9) from the sequence of candidate event times, counting times, efficiencies, and exposure time; the right side of this equation is simply the total production rate. This rate is given in the last column of Table I for 12 Los Alamos runs. It is the combination of the background rate and the neutrino-induced rate.

### D. Background production of $^{127}$Xe

$^{127}$Xe can be produced in the LAMPF detector by means other than neutrinos. The dominant background producing

reaction is $^{127}$I$(p,n)^{127}$Xe, which can be initiated in several ways: (1) cosmic rays can cause the photonuclear evaporation of a proton directly from a nucleus in the target; (2) neutrons from local sources can undergo $(n,p)$ reactions in the water, liberating protons; or (3) $\alpha$ particles from the decay of unstable nuclei in the target can transfer energy to the proton through $(\alpha,p)$ scattering.

As indicated in Table I, five data runs were made with the proton beam off, which directly measured the production rate from the sum of cosmic rays and internal radioactivity. The combined ML analysis of these five runs gave a beam-off production rate $p_{bkgNB} = 46.9 \pm 1.4$ $^{127}$Xe/d.

The background rate from beam-associated neutrons was determined by a combination of measurement and calculation, as will now be described. The neutron flux produced by the proton beam in the detector room was measured during the initial phase of a previous experiment [16]. On the basis of that measurement and the Los Alamos neutron propagation code LAHET [17], the differential spectrum $N$ of neutrons of energy $E$ reaching the experimental area was expected to be $dN/dE = \phi_o E^{-1.8}$, where $\phi_o = 1.5$ neutrons/(MeV mA cm$^2$ d) [18]. Changes in $\phi_o$ of as much as 20% could be expected for the different beamstop configurations used during the measurements. Yet since the neutron-induced background is negligible, as shown below, we can ignore this complication. This neutron flux from the beam was moderated by the 60 cm of water shielding around the tank.

With the flux and spectrum of neutrons that impinge on the NaI solution known, a Monte Carlo program was developed to calculate the $^{127}$Xe production probability from this background source. This program generated a neutron which scattered from a proton in the detector solution. The proton was then followed as it lost energy by ionization and/or reacted via $^{127}$I$(p,n)^{127}$Xe. A check of this program was made by inserting a PuBe neutron source in the middle of a spherical vessel which contained NaI solution. The measured rate was $670 \pm 10$ Xe/h, in good agreement with the predicted rate of $650 \pm 70$ Xe/h.

Convoluting the known neutron flux with the Monte Carlo calculated $^{127}$Xe production probability, it was predicted that $0.20 \pm 0.02$ $^{127}$Xe/(d mA) would be produced by neutrons in the water-shielded tank. The total beam-associated production rate, as shown below, is $\approx 10$ $^{127}$Xe/(d mA). The contribution of neutrons to the total production rate by this calculation is, therefore, 2% of the total beam-associated signal, sufficiently small to be neglected.

### E. Cross section determination

Figure 6 is a plot of total production rate versus effective neutrino flux $F$ for the 12 Los Alamos runs given in Table I. By Eq. (9), the intercept at zero neutrino flux is the beam-off, background production rate, $p_{bkgNB}$ and the slope is the flux-shape weighted cross section $\overline{\sigma}_\nu$ times $N_I$. A least squares fit to the data points in Fig. 6, weighted by their statistical errors, gave a fitted slope of $0.179 \pm 0.057$ (stat) [$^{127}$Xe/d] / [$10^6$ neutrinos/(cm$^2$ s)]. Since the number of iodine target atoms was $7.31 \times 10^{27}$, the total cross section is thus $\overline{\sigma} = [2.84 \pm 0.91 \text{ (stat)}] \times 10^{-40}$ cm$^2$.

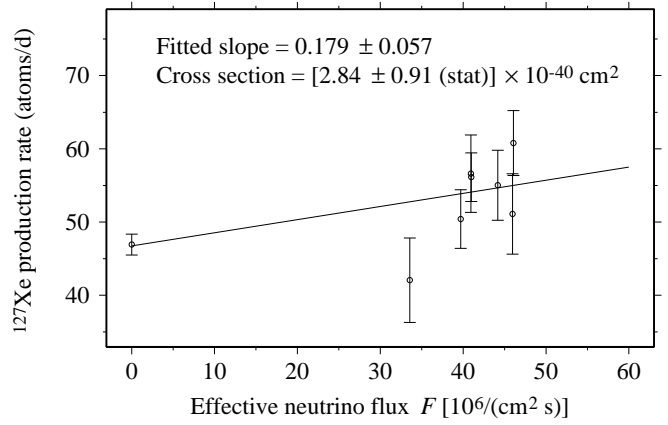

FIG. 6: Production rate of $^{127}$Xe vs neutrino flux in the LAMPF detector tank. The error bars are statistical. The beam-off background point is the maximum-likelihood combination of the five individual background runs in Table I. The straight line is a weighted least-squares fit to the data. The chi-squared for the fit is 7.0 which, with six degrees of freedom, has a probability of 32%.

The uncertainty in this cross section comes from several sources. The statistical uncertainty for each run was determined in the ML analysis and the average for the 12 runs was 8.2%. When translated to the error on the slope, the statistical uncertainty in the cross section became 32%. This large value is a consequence of the poor signal-to-noise ratio of about 1/6. Our experiment would have benefited greatly if the shielding from cosmic rays had been significantly thicker.

The dominant source of systematic error was the 7.3% uncertainty in the Monte Carlo prediction of the neutrino flux produced in the beamstop. Other errors which affected the neutrino flux were the uncertainties in the spatial extent of the source region and the tank-to-source distance. Combined in quadrature these gave a systematic uncertainty in the flux at the tank containing the I solution of 8.0%.

Many of the systematic errors during counting were correlated for different runs. For example, the NaI counter that surrounded the proportional counters was the same for many runs, so the systematic uncertainty in its efficiency was the same for one run as for the final result. Similarly, since the same counters were used for several runs, there was a strong correlation of counter efficiency systematic error. The combination of these uncertainties gave a systematic uncertainty during counting of 3.7%.

Adding these effects in quadrature, the total systematic error was 8.8% or $0.25 \times 10^{-40}$ cm$^2$, considerably smaller than the statistical uncertainty. Because of the negligibly small contribution to the production rate from beam-associated neutrons, as shown in Sec. III D, our final result for the cross section is thus $[2.84 \pm 0.91 \text{ (stat)} \pm 0.25 \text{ (syst)}] \times 10^{-40}$ cm$^2$.

## IV. SUMMARY AND CONCLUSIONS

The cross section for the conversion of $^{127}$I to $^{127}$Xe by electron neutrinos from the decay of stopped muons has been measured to be $[2.84 \pm 0.91 \text{ (stat)} \pm 0.25 \text{ (syst)}] \times 10^{-40}$ cm$^2$. The two existing theoretical predictions for this cross section are $(2.1\text{--}3.1) \times 10^{-40}$ cm$^2$ by Engel et al. [9] and $4.2 \times 10^{-40}$ cm$^2$ by Kosmas and Oset [19]. Our result is in reasonable agreement with both of these predictions. This suggests that the prediction of Engel et al. of $3.3 \times 10^{-42}$ cm$^2$ for the $^8$B cross section (equivalent to 18 SNU) can be used as a reasonable estimate of the $^8$B capture rate to be expected in an I solar neutrino experiment.

It should be noted that the 50% range in the predictions by Engel et al. is due to uncertainty in the strength of the spin-operator (Gamow-Teller) transitions; a similar uncertainty affects calculations of the $^{13}$C$(\nu_e, e^-)$X reaction [20]. More accurate measurements of neutrino reactions on $^{127}$I and $^{13}$C might then serve to measure the Gamow-Teller quenching.

The other nuclei for which a reaction cross section of the inverse $\beta$ decay type has been published are $^2$H and $^{12}$C. The reaction $^2$H$(\nu_e, e^-)pp$, measured in an experiment at LAMPF [21], gave a cross section averaged over the Michel spectrum for $\nu_e$ of $(5.2 \pm 1.8) \times 10^{-40}$ cm$^2$, in good agreement with calculations. There are three measurements of the electron neutrino-induced transition $^{12}$C$(\nu_e, e^-)^{12}$N$_{g.s.}$, two done at LAMPF and the other by the KARMEN detector at ISIS at the Rutherford Laboratory. These measurements in units of $10^{-42}$ cm$^2$ are, respectively, $10.5 \pm 1.0$ (stat) $\pm 1.0$ (syst) [22], $8.8 \pm 0.3$ (stat) $\pm 0.9$ (syst) [23], and $9.3 \pm 0.4$ (stat) $\pm 0.9$ (syst) [24]. The most recent theoretical prediction for this reaction, $8.1 \times 10^{-42}$ cm$^2$ [25], is in good agreement with these measurements. It is reassuring to note that the only two neutrino-nucleus cross section measurements that span the $^8$B range have reasonable agreement with theoretical predictions.

There have also been two measurements of electron neutrino-nucleus cross sections in the $^7$Be range, both using MCi sources of $^{51}$Cr and both measuring the transition rate of the reaction $^{71}$Ga$(\nu_e, e^-)^{71}$Ge, where the states populated in $^{71}$Ge were the g.s. and excited states at 175 keV (5/2$^-$) and at 500 keV (3/2$^-$). The ratio of measured to predicted cross sections was $0.95 \pm 0.12$ (SAGE [26]) and $0.92 \pm 0.08$ (GALLEX [27]), indicating that this technique can be successfully carried out and provides very good agreement between prediction and observation.

The above experience with both $^{12}$C and $^{71}$Ga suggests that the response of an $^{127}$I detector to solar electron neutrinos over the entire energetically allowed spectrum can be directly determined. The remaining steps in the $^{127}$I electron neutrino calibration might include determination of the $^7$Be electron neutrino cross section with a MCi $^{37}$Ar source (814-keV electron neutrinos) and determination of the $^8$B electron neutrino interaction rate by a repetition of the LAMPF measurement with an electronic version of the NaI detector that can measure the $^{127}$I$(\nu_e, e^-)^{127}$Xe cross section as a function of energy.

### Acknowledgments

We gratefully thank W. Haxton and P. Vogel for helpful discussions, J. Klein and D. Storm for making the $^{127}$Xe used in counter calibration, and the Pennsylvania Turnpike Authority for providing a deep location for counter measurement. This work was supported in part by the Physics and Astronomical Sciences Divisions of the National Science Foundation and by the Nuclear Physics Division of the Department of Energy.